\documentclass[sigconf]{acmart}

\usepackage{booktabs} 
\usepackage{enumitem}
\usepackage{subcaption}
\usepackage{dblfloatfix} 
\usepackage[utf8]{inputenc}
\usepackage{balance}

\setcopyright{none}
\usepackage{xcolor}

\acmDOI{}

\acmISBN{}

\acmConference[arXive]{arXive}{April 2018}{https://arxiv.org} 
\acmYear{2018}
\copyrightyear{2018}
\acmPrice{}

\begin{document}
\title[Neuroscientific User Models]{Neuroscientific User Models:\\ The Source of Uncertain User Feedback and Potentials for Improving Recommendation and Personalisation}

\author{Kevin Jasberg}
\affiliation{%
  \institution{Web Science Group -- Heinrich-Heine-University}
  \city{Duesseldorf} 
  \state{Germany} 
  \postcode{45225}
}
\email{kevin.jasberg@uni-duesseldorf.de}

\author{Sergej Sizov}
\affiliation{%
  \institution{Web Science Group -- Heinrich-Heine-University}
  \city{Duesseldorf} 
  \state{Germany} 
  \postcode{45225}
}
\email{sizov@hhu.de}

\renewcommand{\shortauthors}{}

\begin{abstract}
Recent research revealed a considerable lack of reliability for user feedback when interacting with adaptive systems, often denoted as user noise or human uncertainty. Moreover, this lack of reliability holds striking impacts for the assessment of adaptive systems and personalisation approaches. 
Whenever research on this topic is done, there is a very strong system-centric view in which user variation is something undesirable and should be modelled with the eye to eliminate. However, the possibilities of extracting additional information were only insufficiently considered so far.

In this contribution we consider the neuroscientific theory of the Bayesian brain in order to develop novel user models with the power of turning the variability of user behaviour into additional information for improving recommendation and personalisation. To this end, we first introduce an adaptive model in which populations of neurons provide an estimation for a feedback to be submitted. Subsequently, we present various decoder functions with which neuronal activity can be translated into quantitative decisions. The interplay of cognition model and decoder functions lead to different model-based properties of decision-making. This will help to associate users to different clusters on the basis of their individual neural characteristics and thinking patterns.  By means of user experiments and simulations, we show that this information can be used to improve the standard collaborative filtering.
\end{abstract}

\maketitle

\section{Introduction}
Personalisation and recommendation have become indispensable in most systems nowadays and the trend still continues to grow in that direction. During the last decade, the growth of interactions continuously supported innovations in a data-driven fashion. This is advantageous as we need to understand a user along with his preferences, peculiarities and behaviour to adapt recommendation and personalisation in order to provide an appealing user experience. This is done by inventive user models and by injecting information into modern personalisation engines based on techniques of machine learning, but the bedrock of such efforts is a thorough knowledge about the user, either by observation (implicit knowledge) or by questioning (explicit knowledge). 

\begin{figure}[t]
    \centering
    \begin{subfigure}{0.235\textwidth}
        \includegraphics[width=\textwidth]{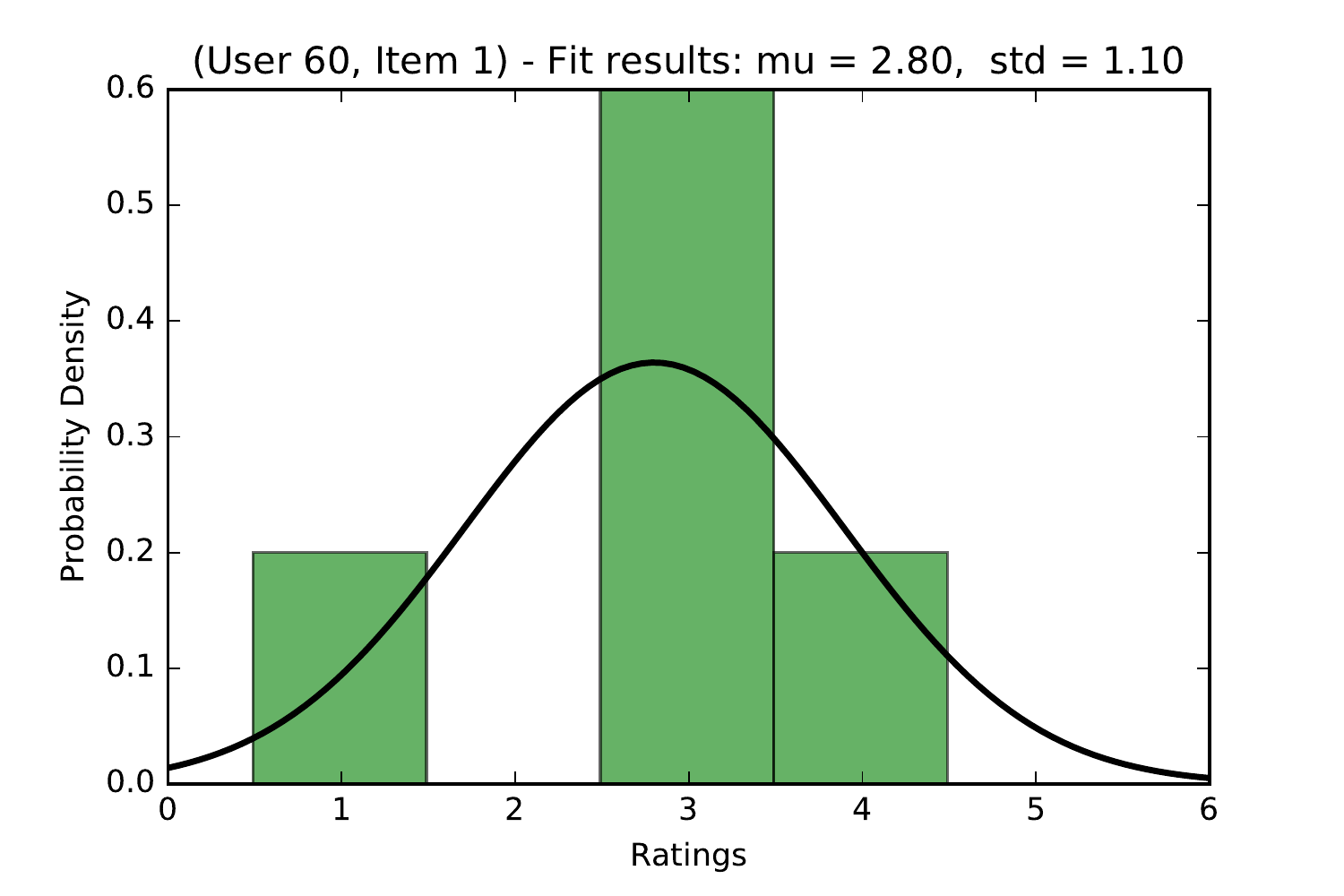}
    \end{subfigure}
     \hfill
    \begin{subfigure}{0.235\textwidth}
        \includegraphics[width=\textwidth]{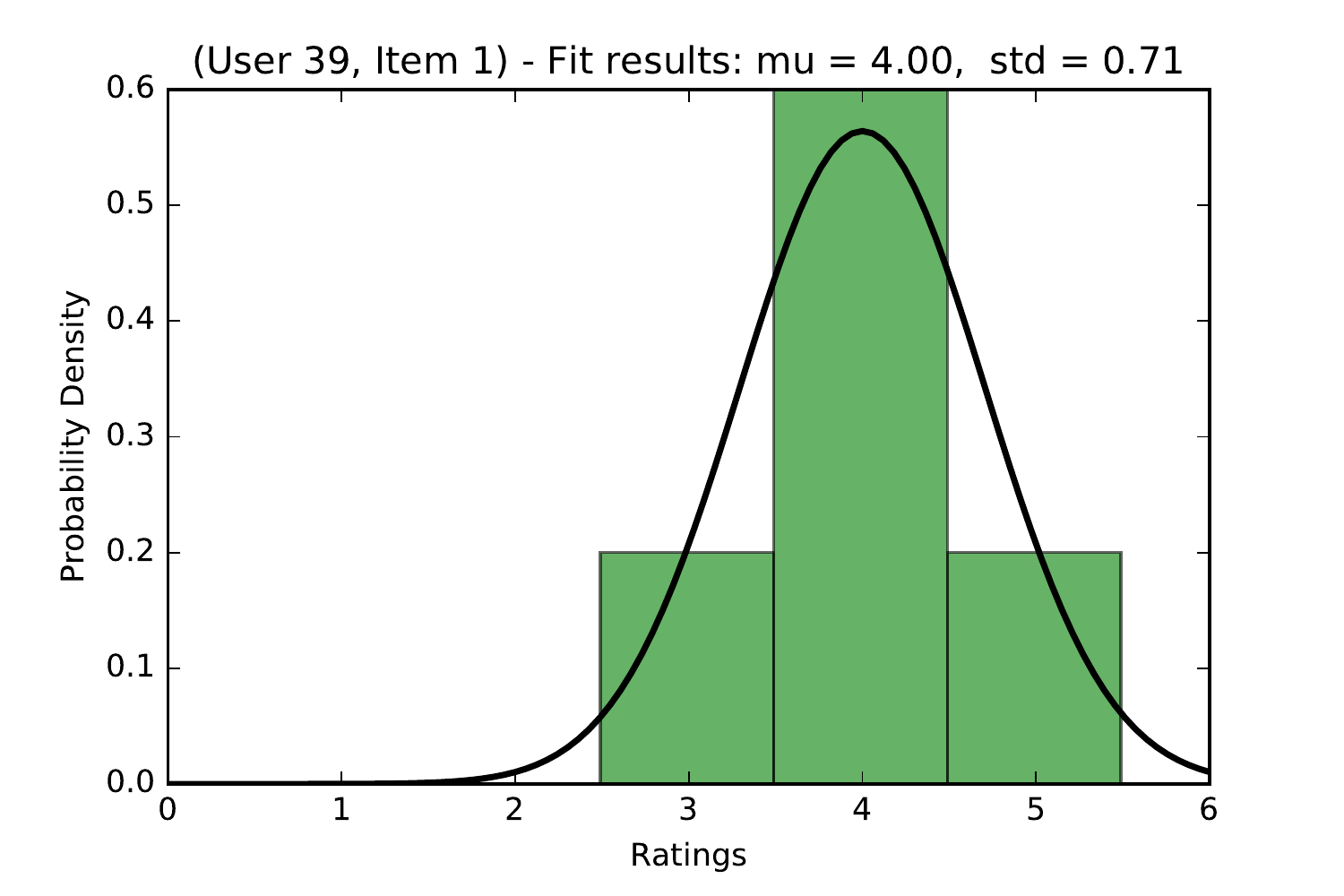}
    \end{subfigure}
         \hfill
    \begin{subfigure}{0.235\textwidth}
        \includegraphics[width=\textwidth]{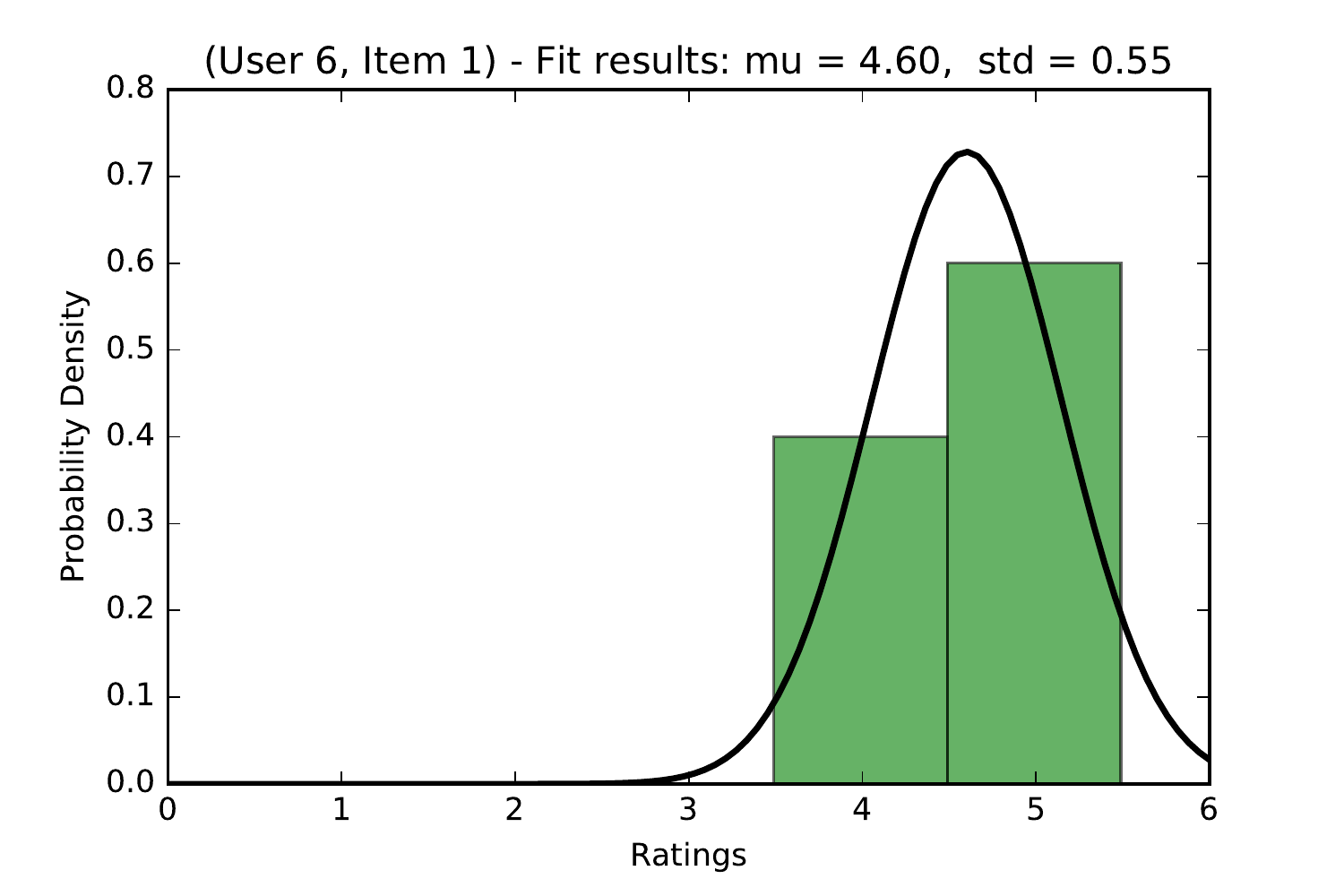}
    \end{subfigure}
         \hfill
    \begin{subfigure}{0.235\textwidth}
        \includegraphics[width=\textwidth]{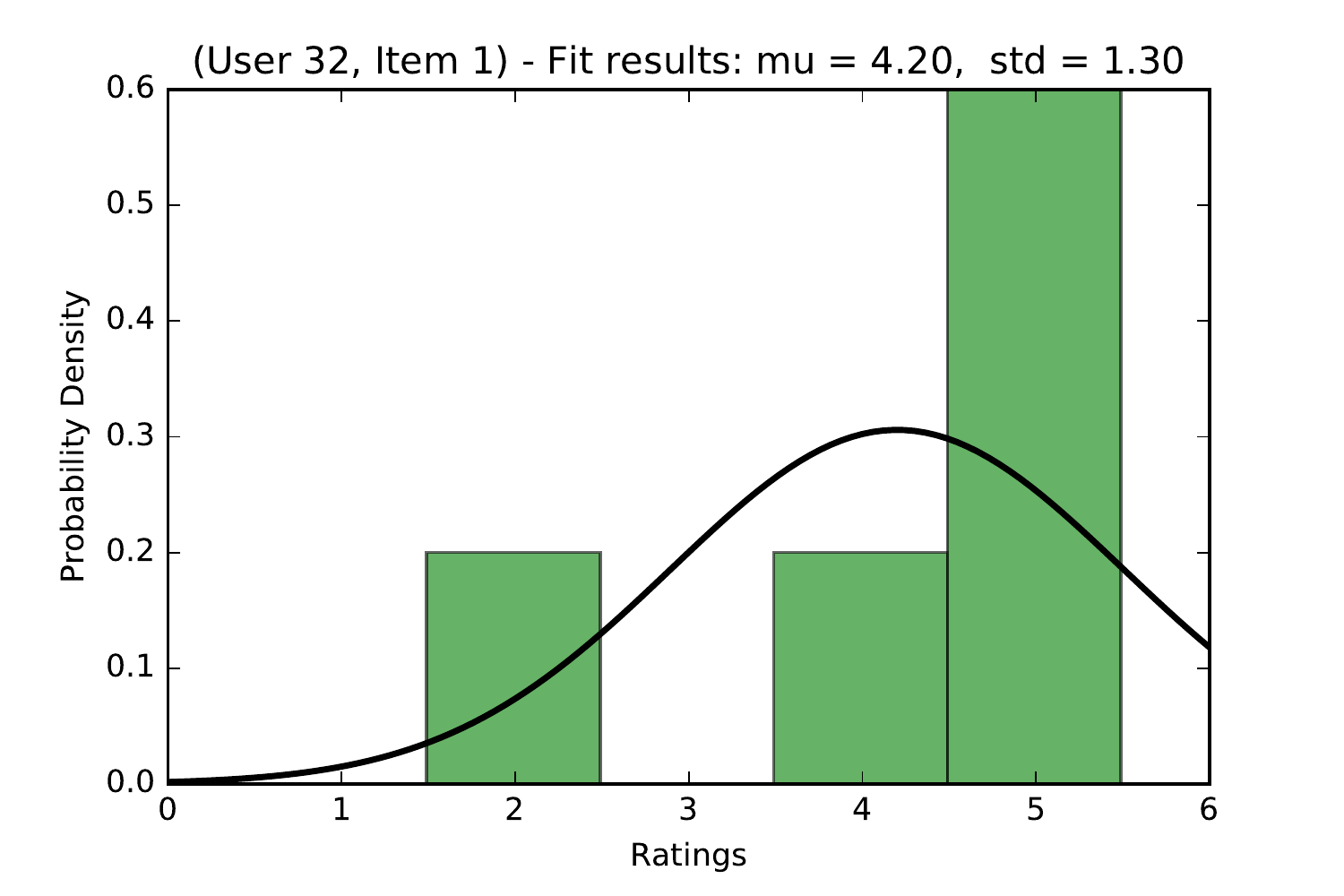}
    \end{subfigure}
    \vspace{-3ex}
    \caption{Visualisation of uncertain user responses for a repeated feedback task.}
    \vspace{-3ex}
    \label{fig:UserResponsesExample}
\end{figure}

The strong dependence on user-generated data is curse and blessing at the same time, because the fundamental problem with user feedback is its uncertainty. This means that a considerable fraction of users behave differently in the same context or decide otherwise if a decision task has to be repeated. This phenomenon is often denoted as user noise (or human uncertainty in recent research) and gives user data the nature of a random variable.
As an introductory example we consider the repeated rating of film trailers in a short temporal interval (granting same emotional and cognitive context) with a sufficient number of distractors between rating repetitions.
Figure \ref{fig:UserResponsesExample} shows the different ratings of four users of this experiment, which will be described in more detail in forthcoming sections. It becomes clear that this user feedback is scattering around a central tendency and hence supports the assumption of a random variable. This feature has recently been in the focus of some research that presented both, induced problems in the evaluation of adaptive systems as well as attempts for possible solution strategies.

The commonality of all these contributions in the field of user modelling, adaptation and personalisation is that there is a very strong system-centric view in which user variation is something undesirable and should be modelled with the eye to eliminate. All developed solutions more or less simply try to ignore uncertain data (which obviously leads to results with less uncertainty as well) but they are by no means satisfactory and thus we have to ask whether this controversial view amidst a large fraction of researches is yet worthwhile. In this contribution, we want to introduce a new and diametrically opposed paradigm in which we consider uncertainty no more as a mistake or dysfunction with destructive side effects, but rather as an opportunity for gathering additional information. 
Such an undertaking sensibly has to start with the measurement of user feedback along with its uncertainty and by transmission of this data to a user model, which maps uncertainty into an information space in order to successively supplement a user profile. 
Since the measurement of user noise or human uncertainty has already been a subject of research, we confine ourselves to the development of a novel user model with special sensitivity for response uncertainty. This naturally leads to the following research questions:
\begin{enumerate}
\item What does a possible model look like that considers human decision variability and maps it into the highest possible concentration of additional information?
\item How can this information be integrated into existing recommender systems and personalisation engines?
\item What are the final benefits of this particular user model and what are the benefits of this novel paradigm in general?
\end{enumerate}

\section{Related Work}

\textbf{Recommender Systems and Assessment.}
A lot of research about recommendation and personalisation produced a variety of techniques and approaches \cite{Jannach, Handbook}. For the comparative assessment, different metrics are used to determine the prediction accuracy, such as the root mean squared error (RMSE), the mean absolute error (MAE), along with many others \cite{Herlocker, Bobadilla}. In our contribution, we internalise existing criticism about a lack of understanding human beings in the process of system design \cite{McNee,Knijnenburg} and develop a user model that is close to the current way of looking at the functionality of the human brain.\\[.125ex]

\textbf{Dealing with Uncertainties.}
The relevance of our contribution arises from the fact that the unavoidable human uncertainty sometimes has a vast influence on the evaluation of different prediction algorithms \cite{LikeLikeNot, noise2}. The idea of uncertainty is not only related to recommender systems but also to measuring sciences such as metrology. Recently, a paradigm shift was initiated on the basis of a so far incomplete theory of error \cite{Grabe, Buffler}. In consequence, measured properties are currently modelled by probability density functions and quantities calculated therefrom are then assigned a distribution by means of a convolution of their argument densities. This model is described in \cite{GUM}. 
We transfer this perspective to user feedback by considering it as a single draw from an underlying distribution. 
This provides us with a probabilistic reference which we can use to verify the predictions based on our own user model.\\[.125ex]

\textbf{The Idea of Human Uncertainty.} 
The idea of underlying distributions for user feedback is not far-fetched since the complexity of human perception and cognition can successfully be addressed by means of latent distributions \cite{delia}. We adopt the idea of modelling user uncertainty by means of individual Gaussians for constructing our individual response models and thus follow the argumentation in latest research of neuroscience and metrology \cite{GUMsupp1, Pouget}.
Probabilistic modelling of cognition processes is also quite common to the field of computational neuroscience. In particular, aspects of human decision-making can be stated as problems of probabilistic inference \cite{FristonNature}
(often referred to as the ``Bayesian Brain'' paradigm). At any time when a decision has to be made, one has to consider a variety of yet unknown states of the world which are most relevant for the decision process itself. According to \cite{Friston}, each of these states are unconsciously estimated by a population of neurons (agency) and thus being made accessible to the brain. In doing so, there is evidence that each agent provides a probability density over possible values of such a state of the world (probabilistic populaton codes) and thus also accounts for its uncertainty \cite{Pouget}. However, these estimations slightly differ in each cognition trial due to the volatile concentration of released neurotransmitters, impacting the spiking habits of downstream neurons (neural noise) \cite{NoisyNS, Pouget}. In other words, human decisions can be seen as uncertain quantities by nature of the underlying cognition mechanisms.
In this paper, we adopt the theory of noisy probabilistic population codes (nPPCs) and use them to construct a user model that can naturally represent and explain response uncertainty with neural noise, mapping this uncertainty to specific neural parameters.\\[.125ex]

\textbf{Human Uncertainty in Computer Science.}
User noise or human uncertainty has been mentioned before in computer science.
The first reference in the context of user feedback came along with a study on the applicability of thinking strategies to product recommendations, where reliability problems were registered for repeated ratings \cite{Hill}.
The authors, like \cite{Herlocker} later on, have already speculated on its impact on adaptive systems' accuracy.
This assumption was later confirmed when uncertainty in user ratings (measured by re-rating) and its impact on the RMSE was demonstrated \cite{LikeLikeNot,RateAgain}. However, this impact was a mere deterioration by means of a specific metric. A more sophisticated analysis is provided by \cite{MagicBarrier}, where it could be demonstrated that human uncertainty leads to an offset in a specific metric (magic barrier). This approach was later expanded by \cite{JasRecSys} and it was shown that this barrier has some uncertainty itself, i.e.  even if RMSE scores are not below this barrier, they could already be completely random. Additional contributions also revealed that each accuracy metric considering human feedback is naturally biased and that possible rankings built upon these metric scores are subject to probabilities of error \cite{JasSAC}.

To solve this problem, some strategies have been proposed over the years, just like a pre-processing step that deletes highly deviant values and replaces them by artificial values closer to the mean of a re-rating \cite{RateAgain}. Another approach is to provide  model-based predictions with uncertainty as well, so that the uncertainties of a rating and a predictor eliminate each other when calculating their difference \cite{OrdRec}. Yet another possibility is to compute metrics only with deviations that differ widely from a given predictor and hence can not be explained by human uncertainty \cite{JasWISE}. With our contribution, we want to move away from the paradigm of extinction and present a way in which uncertainty can be sensibly used to generate benefits.

\section{Theory, Models and Applications}
\subsection*{The Single Neuron Model}
The response of a single neuron to a stimulus is limited to transmission of electric impulses (spiking) and since each neuron has only got two states of activation, theories of neural coding assume that information is encoded by the spiking frequency (rate) \cite{BayesianBrain}. The functional relationship between responses $r$ of a neuron and the characteristics $s\in S\subset\mathbb{R}$ of a stimulus is given by the so-called tuning curve $r=f(s)$. 
Besides irregular shapes, tuning curves have frequently been measured to be bell-shaped or sigmoid-shaped respectively.
Each tuning curve maximises for a particular value $p:=\operatorname{argmax} f$, denoted as the preferred stimulus.
For bell-shaped tuning curves, $f\colon S \to \mathbb{R}$ can be modelled as
\begin{equation}
f_{p}(s) :=g\cdot h(p,w)(s) + o,
\end{equation}
where the shape emerges from the Gaussian density function $h$ with mean $p$ and standard deviation $w\in\mathbb{R}^{>0}$ (tuning curve width). The additional components $g\in\mathbb{R}^{>0}$ and $o\in\mathbb{R}^{>0}$ represent a frequency gain and offset respectively.

When measuring tuning curves in reality, one will find that they are somewhat noisy and that even one and the same stimulus never leads to the same response. This fluctuations can be explained by the so-called neural noise \cite{NoisyNS}.
Neuronal responses must therefore be seen as random variables $R$ rather than fixed values determined by tuning curves. It has been found that $R\!\sim\!\operatorname{Poi}(\lambda)$ follows a Poisson distribution with expectation $\lambda=f(s)$ \cite{Pouget,Tolhurst}.

\subsection*{Probabilistic Population Codes}
\begin{figure}[t]
    \centering
    \includegraphics[width=\linewidth]{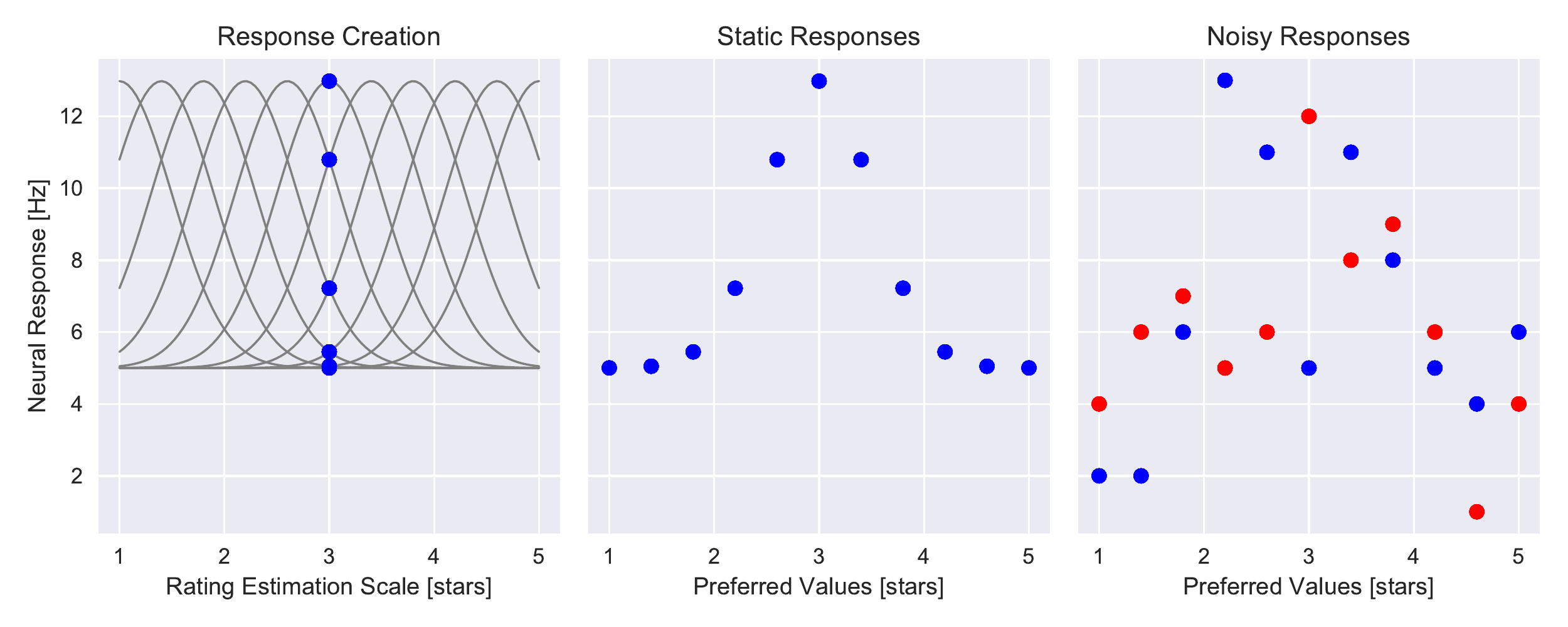}
    \caption{Genesis of noisy population responses demonstrating the alteration for each cognition trial (red and blue).}
    \label{fig:NoisyResponseOrigin}
	\vspace{-2ex}
\end{figure}

We now consider a population of $n$ neurons, all with the same tuning curve type with (almost) the same neural parameters. The only difference is in the preferred values $p_j$, which are equidistantly spread across the range of possible stimuli (estimation scale).
Mathematically we realise this by considering the real continuous set $S\subset\mathbb{R}$ and a sequence $(p_j)_{j=1,\ldots,n}$ being an equidistant discretisation of $S$. All parameters determining the population size $n$, the shape of all tuning curves as well as the assumed stimulus $s\in S$ are summarised in a vector $\xi = (n,g,w,o,s)$ which we will refer to as the cognition vector in the following.
Given a particular fixed $s$ (which is formed from unknown underlying cognitions), each neuron of this population will respond according to its specific tuning curve and interference due to neural noise. Therefore, a response $r_j$ of the $j$-th neuron must be seen as a realisation of the random variable $R_j\sim \operatorname{Poi}( f_{p_j}(s) )$. In order to keep in mind, that these responses are always dependent on the parameters of the cognition vector, we henceforth use the notation $r_j(\xi)$ as realisation of $R_j(\xi)$. The response of the entire population is formed by the response of each neuron and so we denote the $n$-dimensional random variable 
\begin{equation}
\mathcal{R}(\xi) := \left(R_1(\xi) \, ,\ldots , \, R_n(\xi) \right)
\end{equation}
as the population response for a given $\xi$ with realisation $\varrho(\xi) = \left(r_1(\xi) \, , \ldots , \, r_n(\xi) \right)$. 
This theory of the origin of noisy population responses is illustrated in Fig. \ref{fig:NoisyResponseOrigin}.
In this example, we used $\xi= (11,10,0.5,5,3)$ as the cognition vector , i.e. we consider $n=11$ neurons that respond to the assumed stimulus (in this case: cognition result) of $s=3$ stars where each tuning curve has the offset $o=5\,\text{Hz}$, the width $w=1\,\text{Hz}$ and the gain $g=7$. In the left picture we can see the individual tuning curves, which are distributed equidistantly over the possible range of a rating scale with five stars. For $s=3$ stars, the responses of each neuron can be fetched from its tuning curve. For a better representation of the population response, it has become a standard to plot the individual responses against the corresponding preferred values, which can be seen in the middle picture. These are the theoretical (static) responses without consideration of neural noise. To add this neural noise, each static response $r^\text{static}_j(\xi)$ is replaced by the draw of a random number from the Poisson distribution with parameter $\lambda = r^\text{static}_j(\xi)$.
This can be seen in the right subfigure. We additionally repeated this sampling once, i.e. the blue and red dots in each case represent a noisy population response and it is obvious that these population responses differ not only from the theoretical reference but also very much from each other. At this point, we see that the same cognition (represented by $\xi$) leads to different neural activities on each pass, and that the estimation of a quantity (e.g. product rating) or state of the world is thereby given a natural uncertainty. 

\subsection*{Decoder Functions}
What we have learned so far is the internal basic cognitive model that allows different neuronal activity for a population of neurons to encode one and the same state of the world. By means of sensory perception, this model can be seen as the a translation of outside reality into inside representation of the external world. By means of cognition, however, this model provides the translation from a cognition black box into internal and measurable representations of thoughts and thinking patterns. 
The main question that arises at this point is: How does the human brain translates population activity into estimations for a state of the world or a cognition respectively. Theories assume the use of so-called decoder functions.
Mathematically, a decoder function is a mapping $\varphi\colon\mathbb{R}^n\to S$ from population activity onto the estimation scale for a stimulus or cognition. This means that for a particular user-item-pair $(u,i)$, we can obtain an estimation of a single feedback submission directly from the realisation of a population response, i.e. $f_{u,i} = \varphi(\varrho(\xi))$. Hence, noisy user feedback $\mathcal{F}_{u,i}$ can be represented as a random variable given as
\begin{equation}
\mathcal{F}_{u,i} = (\varphi\circ\mathcal{R}) (\xi).
\label{eq:FeedbackOrigin}
\end{equation}
In neuroscience literature, there are several decoders that have been suggested and frequently used so far \cite{PougetDecoder}. We will give a brief overview of the most frequently discussed decoder functions and will relate them directly to the context of user feedback.

\begin{figure}[t]
    \centering
    \includegraphics[width=\linewidth]{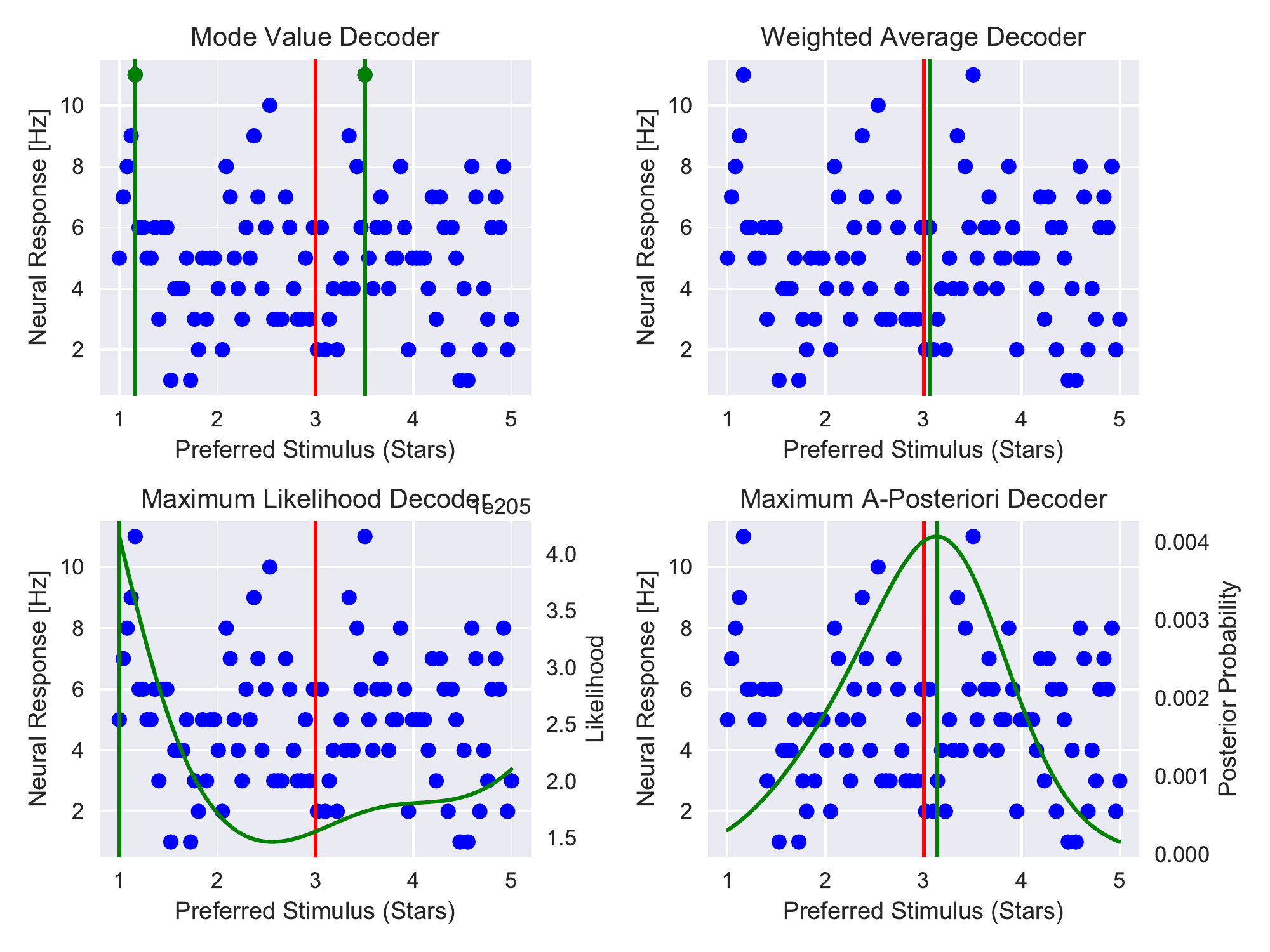}
    \caption{Visualisation of decoder functions on a population response for a given $\xi=(100,1,1,5,3)$. 
    The red and green lines show the true and the decoded stimulus.}
    \label{fig:FourDecoders}
	\vspace{-2ex}
\end{figure}

\paragraph*{Mode Value Decoder}
Due to the construction of tuning curves, the MVD assumes that it is exactly the neuron with maximum spiking frequency that is most likely to be addressed by the stimulus or the state of the world. The decoder function is thus given as
\begin{equation}
\varphi_{MVD}\colon \varrho(\xi) \mapsto \operatorname*{argmax}_{p_j\in S} \{r_1(\xi),\ldots, r_n(\xi) \}.
\end{equation}
Figure \ref{fig:FourDecoders} depicts a population response for a 3-star-decision (red line) together with possible estimators (green lines) for this decision. This decoder is very prone to neural noise and its estimators are subject to a great ambiguity which, however, diminishes for higher frequencies in neural responses.

\paragraph*{Weighted Average Decoder} The WAD accounts for all responses by setting the specific frequency as a weight to the corresponding preferred value  and considers its contribution to the total response. In mathematical terms, the WAD is given by
\begin{equation}
\varphi_{WAD}\colon \varrho(\xi) \mapsto 
\frac{\sum_{j=1}^n r_j(\xi)\cdot p_j}{\sum_{j=1}^n r_j(\xi)}.
\end{equation}
As to see in Fig. \ref{fig:FourDecoders}, this decoder function does not produce ambiguous estimators and is very stable against neural noise. 

\paragraph*{Maximum Likelihood Decoder}
For a given population response, the MLD chooses the estimator $\hat s$ with a view to maximise the corresponding likelihood function, i.e.
\begin{equation}
\varphi_{MLD}\colon \varrho(\xi) \mapsto  \operatorname*{argmax}_{s\in S} \,\,
P( \varrho(\xi) | s),
\end{equation}
where the likelihood itself is given by the i.i.d.-assumption together with the Poisson probability mass function
 \begin{eqnarray}
 P(\varrho(\xi)|s) &=& P(r_1(\xi),\ldots , r_1(\xi)|s) \nonumber\\
 &=& \prod_{j=1}^n \frac{f_{p_j}(s)^{r_j(\xi)}}{r_j(\xi)!} 
         \exp\left(-f_{p_j}(s)\right).
 \end{eqnarray}
In Fig. \ref{fig:FourDecoders} we see the likelihood function (green curve) for the particular population response together with the MLE estimator (green line). The MLD is the first decoder that explicitly accounts for neural noise through the Poisson probability mass function. 

\paragraph*{Maximum A Posteriori Decoder}
The likelihood can be transformed into a probability function over the stimulus via Bayes' theorem, i.e. 
$P(s|\varrho(\xi)) \propto P(\varrho(\xi)|\hat{s})\cdot P(s)$. $P(s)$ denotes prior belief about the stimulus or the states of world that has been learned through former experiences. The estimator is then chosen so that this posterior is maximised, i.e. 
\begin{equation}
\varphi_{MAD}\colon \varrho(\xi) \mapsto 
\operatorname*{argmax}_{s\in S} \,\,  P(s | \varrho(\xi))
\end{equation}
The MAD is much like the MLD but with less variability since the prior works as a stabiliser. In the example of Fig. \ref{fig:FourDecoders} we arbitrarily used a Gaussian with $\mu=3$ and $\sigma^2=0.75$ as prior belief.
The Bayesian brain theory assumes a prominent role of this decoder function, since each population would then naturally represent a probability density over a stimulus or state of the world which can easily be aggregated with other populations' densities by mere addition. For multiple sensory inputs, this decoder function was proven to be a plausible description for the brain's operating principles \cite{MultiSensory}.

\begin{figure}[b]
    \centering
    \includegraphics[width=\linewidth]{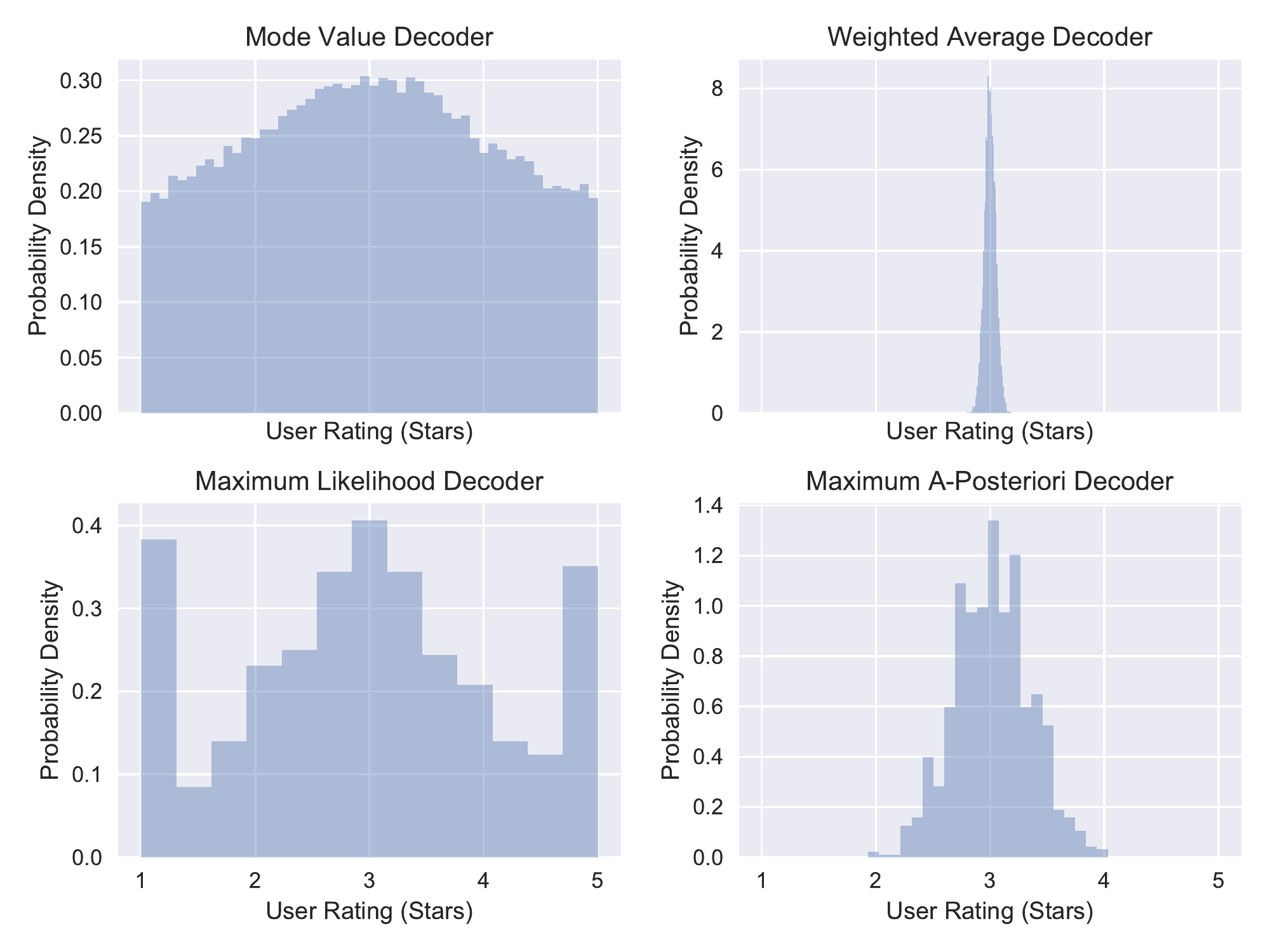}
    \caption{Feedback distributions obtained from different decoder functions for the cognition vector
     $\xi=(100,1,1,5,3)$.}
    \label{fig:FourDists}
	\vspace{-2ex}
\end{figure}

\begin{figure*}
    \centering
    \includegraphics[width=\linewidth]{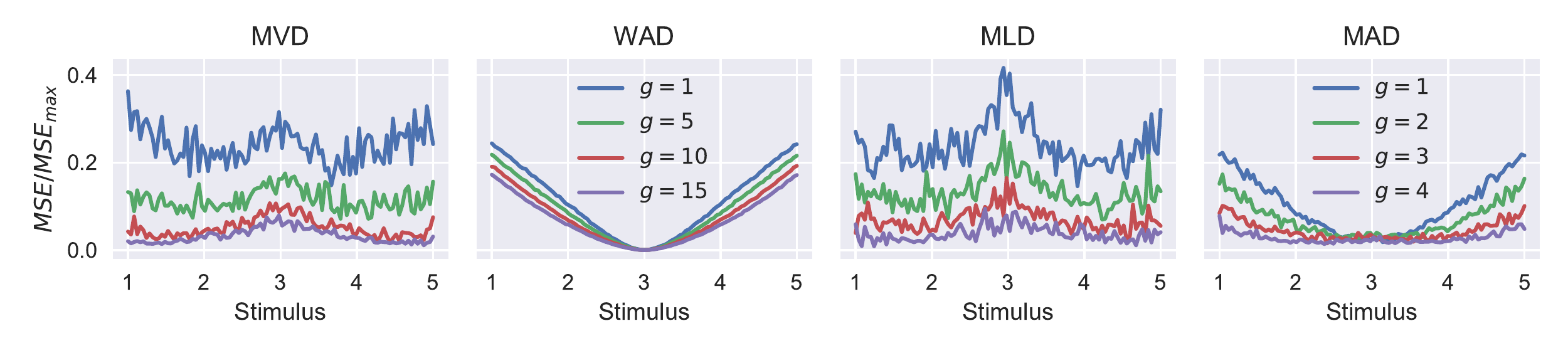}
    \vspace*{-2ex}
    \caption{Reliability Analysis -- Comparing $s$ (input) with possible estimations $\hat{s}$ (output) by means of $MSE/MSE_{max}$ in dependency of the stimulus and tuning curve gain $g$. For the MAD, $\mathcal{N}(s, 0.75)$ was used as informative prior.}
    \label{fig:InOut}
	\vspace{-2ex}
\end{figure*}

\subsection*{Theoretic Model Properties}
As already mentioned, this modelling can explain the genesis of uncertain user feedback $\mathcal{F}_{u,i}$. For the purpose of exemplification, we have computed the resulting feedback distributions for all introduced decoder functions by using Eq. \ref{eq:FeedbackOrigin} for the cognition vector $\xi=(100,1,1,5,3)$. The results are depicted in Fig. \ref{fig:FourDists}. 

Already here, certain properties of this model are clearly visible. 
For the MVD, the vulnerability for neural noise is quite obvious since the corresponding feedback distribution got the largest spread. Even at the boundaries of 1 and 5 stars, there are still high probabilities, hence this distribution is only slightly more informative than a uniform distribution. Using the Bayesian definition of probability (which is interpreted as ones personal confidence), such a user feedback would be provided by users who are not sure about which rating seems appropriate. For the WAD, we notice the robustness to neural noise and the quality of estimation. A user which would utilise this decoder function would surely give constant ratings. Conversely, users with larger uncertainties can probably not be modelled by this decoder. The MLD reveals a remarkable property. Due to the small size of the rating scale $S=[1,5]$, the likelihood's maximum frequently coincides with the scale boundaries. Therefore, this theory might explain the common user behaviour of giving preference to these boundary ratings.
At first glance, the MAD provides the most plausible feedback distributions which seems to strengthen the Bayesian brain theory.

Of course, all of these distributions depend on the neural parameters in the cognition vector, i.e. the ability to decode responses and compute estimators is strongly dependent on many factors. A sensitivity analysis reveals that the strongest dependency is given for the tuning curve gain, which is not surprising as the gain determines the frequency of neural responses and information is neurally encoded by frequencies. A more thorough analysis of the decoding quality is depicted in Fig. \ref{fig:InOut}. By repeated cognition (population response), computed estimators $\hat{s}$ can be compared with the true stimulus $s$ by means of fractions of the maximum mean squared error (MSE). In this case, the MSE has to be divided by its maximum, because a change of $s$ naturally changes the limits of the MSE which biases analyses (e.g. for $s=3$ the MSE can only be $\leq 4$, but for $s=1$ the MSE can be up to $16$).
For all decoders, we see that the estimation quality increases with neural frequency, i.e. the more active the population, the better a cognition can be translated into a numerical estimate.

For the MVD, lower frequencies evoke that the middle of a scale as well as its margins can be estimated slightly worse than the rest. For higher frequencies, it is only the middle of a scale that can be estimated slightly worse. This would inevitably lead to more uncertainty for these values if a rating task is repeated. This decoder thus explains the effect, that  margin ratings are much more reliable. 
For the WAD, we can see the opposite effect. This decoder is suitable for users who give reliable ratings for the middle of a scale. Both decoder functions need high frequencies ($g\geq 10$) in order to work with high quality.
In contrast, the MAD and the MLD are capable of forming  the same quality profile with lower frequencies. This basically means a lower neural energy consumption for a brain while maintaining full functionality (evolutionary advantages). 
Moreover, since the MLD is only a special case of the MAD (with uniform prior), the MAD is the only decoder function forming a variety of quality profiles, for which one would otherwise need two different decoders. Evolutionarily, it is much more reasonable to develop a single mechanism that can be used in all situations than to develop different mechanisms for this task. These arguments can therefore be seen as another indication for the applicability of the Bayesian brain paradigm. This also means that the MAD is again the best candidate for a neuroscientific user model, which is in line with the previous discussion of feedback distributions.

\subsection*{Neuroscientific User Model}
The goal of this user model is to find a specific cognition vector $\xi_{u,i}$ for each user-item-pair $(u,i)$ along with a decoder function $\varphi$, so that the model-based feedback $\hat{\mathcal{F}}_{u,i}$ minimises the difference to the real user feedback $\mathcal{F}_{u,i}$ by means of an arbitrary disparity metric $d$. Mathematically, our user model is given by $\mathcal{F}_{u,i}\equiv (\xi_{u,i},\varphi)$ with
\begin{eqnarray}
(\xi_{u,i},\varphi) 	&:=& 	\operatorname*{argmin}_{(\xi,\varphi)} \,\, 
d\big( \mathcal{F}_{u,i} \, ,\, \hat{\mathcal{F}}_{u,i} \big)   \nonumber \\
&=&	  \operatorname*{argmin}_{(\xi,\varphi)}\,\, d\big( \mathcal{F}_{u,i}\, ,\, (\varphi\circ\mathcal{R}) (\xi_{u,i}) \big). \label{eq:Assign}
\end{eqnarray}
In the case of ambiguity, that is, when several different cognition vectors lead to the same minimum of $d$, we will select the vector $\xi=(n,g,w,o,s)$ that minimises the population energy 
\begin{equation}
E\propto n\cdot(g+o). \label{eq:Energy}
\end{equation} 
This reasoning arises from the fact that a human brain always has to work in an energy-efficient manner and thus is most likely to use the cognition vector, in which all $n$ neurons spike as sparsely as possible. 
The advantage of this model is that each user-item-pair can be mapped into a high-dimensional space that theoretically carries much more information than the consideration of product ratings does alone.

\section{Evaluation and Results}
Although it is clear that this model does not represent the absolute truth about the human brain, the theory of nPPCs has often been confirmed in the context of sensory perception, and in our case this model is (at least in theory) capable of explaining the lacking  reliability of user feedback in same situational contexts. In this section, we will systematically evaluate this theoretic ability and examine how far this model fits the real human uncertainty and how adaptive systems may benefit from it.

\subsection*{Measuring Human Uncertainty (User Study)}
\begin{figure}[t]
    \centering
    \begin{subfigure}{0.49\linewidth}
        \includegraphics[width=\textwidth]{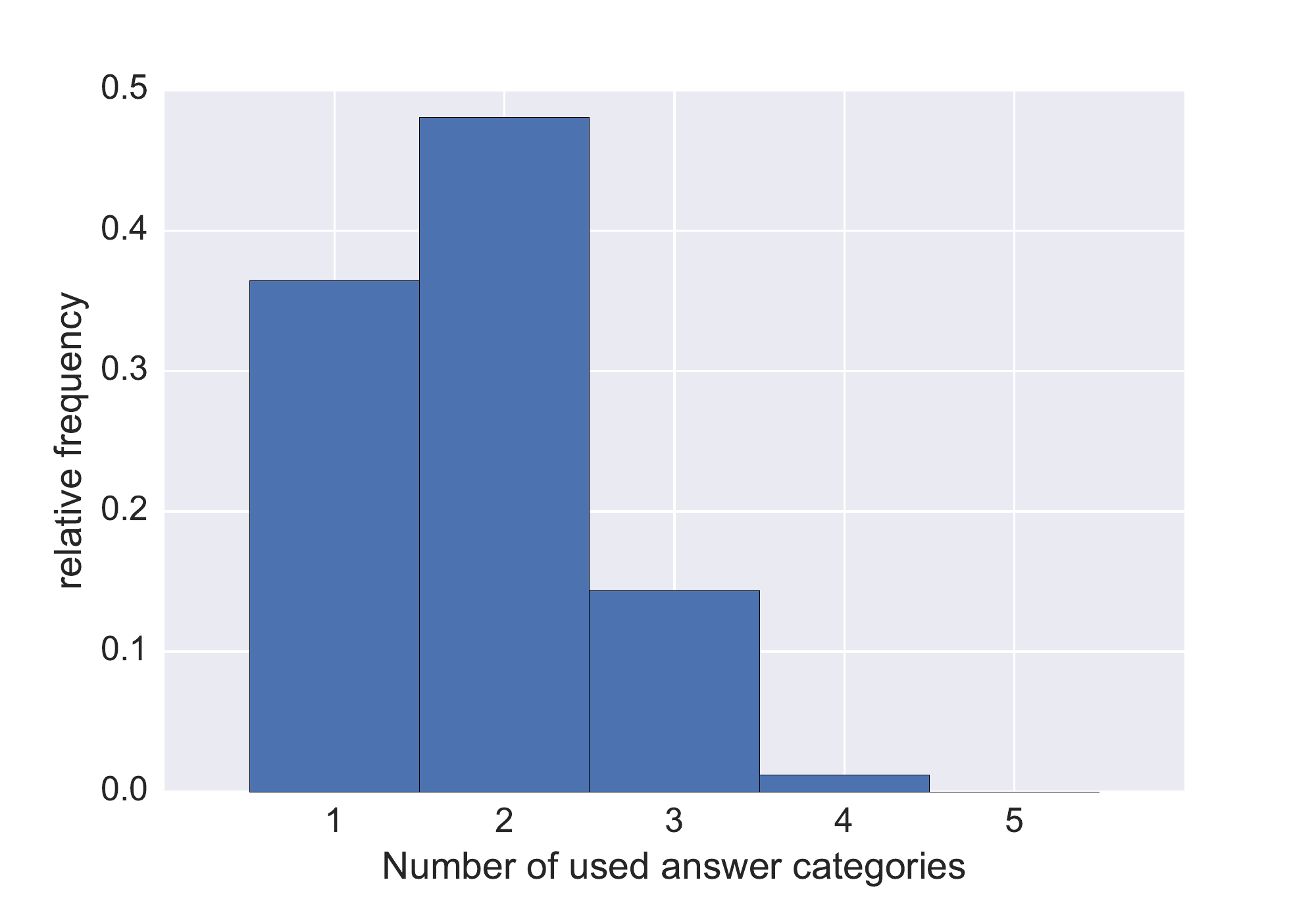}
        \caption{Frequency of used response categories (change in behaviour).}
        \label{fig:ResponseCats}
    \end{subfigure}
    \hfill
    \begin{subfigure}{0.49\linewidth}
        \includegraphics[width=\textwidth]{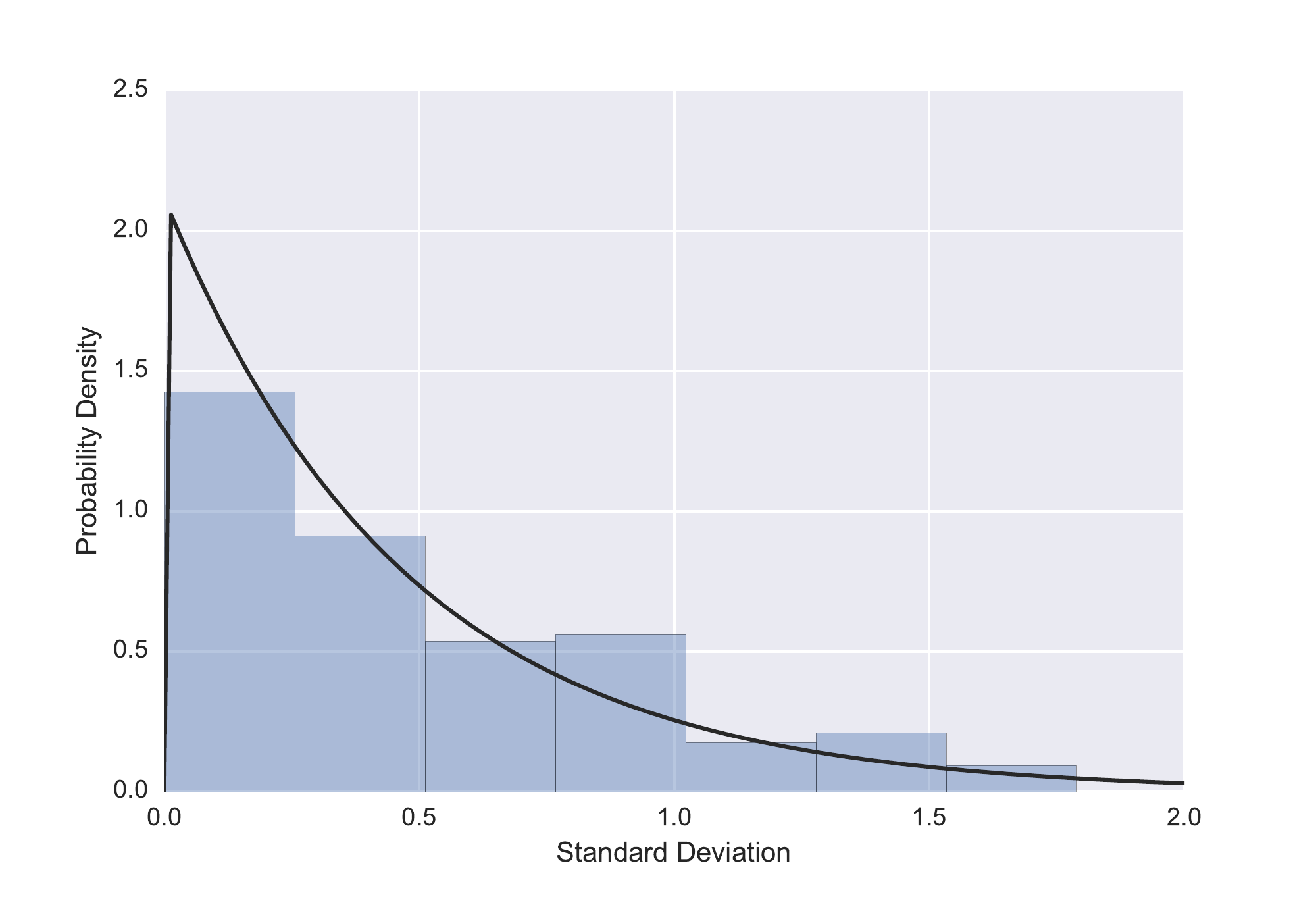}
        \caption{Distribution of Variances for the RETRAIN study}
         \label{fig:DistVar}
    \end{subfigure}
    \vspace{-1ex}
    \caption{Visualisation of human uncertainty found in the RETRAIN user study.}
    \vspace{-2ex}
\end{figure}

Im 2016, we conducted the RETRAIN (Reliability Trailer Rating) study as an online experiment in which 67 participants had watched theatrical trailers of popular movies and television shows and provided ratings in five consecutive repetition trials. User ratings have been recorded for five of ten trailers so that the remaining ones act as distractors, triggering the misinformation effect, i.e. memory is becoming less accurate due to interference from post-event information. The so obtained data set comprises $N = 1\,675$ individual ratings. 
As mentioned before, we discovered that user responses scattered around a central tendency rather than being constant.
From all user ratings, only 35\% manifested a consistent response behaviour, while 50\% gave two different responses on the same item, and 15\% used even three or more different ratings. A detailed breakdown can be found in Fig. \ref{fig:ResponseCats}. The human uncertainty itself is thereby exponentially distributed as to see in Fig. \ref{fig:DistVar}.
In the following, we use this data record\footnote{The data record is available open access at: \textit{link omitted for review}} to fit individual feedback distributions from all ratings that a user has given to the same item. These will then be  compared with our model-based distributions.

\subsection*{User Modelling Quality}
To assign each user-item-pair its own cognition vector and decoder, we compute the model-based feedback 
$\hat{\mathcal{F}}_{u,i}=(\varphi\circ\mathcal{R}) (\xi)$ for each of the four decoder functions $\varphi$ and for each cognition vector $\xi\in N\times G\times W\times O\times S$, where each set
\begin{eqnarray*}
N := \{ 1 , \ldots , 250 \} &  G := \{ 1 , \ldots , 100 \} &  W := \{ 0.1 , \ldots , 2.0 \}  \\
O := \{ 1 , \ldots , 15 \}   & S := \{ 1 , \ldots , 5 \}  		&
\end{eqnarray*}
contains 100 equidistantly distributed values. Altogether, there are $4\cdot 10^9$ combinations to be examined brute force. Subsequently, each $\hat{\mathcal{F}}_{u,i}$ will be compared to the real user feedback $\mathcal{F}_{u,i}$ by means of Eq \ref{eq:Assign}. In doing so, we use two different metrics $d$, one for a discrete evaluation (close to the original data) and another for a continuous evaluation (more accurate, but on basis of assumptions):\\[.25ex]

\begin{description}[leftmargin=6mm]
\item[Cohen's Kappa:] This metric is intended to evaluate inter-rater-reliability and compares the concurrence of two independent classifications with the probability of reaching this agreement by random guessing. This metric is given by the equation $\kappa =(p_{0}-p_{c})/(1-p_{c})$, where $p_{0}$ is the relative agreement of both raters and $p_{c}$ denotes the chance of a random agreement. Its utilisation presupposes discrete finite classes which are given by the discrete rating scale of the RETRAIN study.

To compute $p_{0}$ for each cognition vector and decoder, we draw five model-based estimators $\varphi(\varrho(\xi))$ (rounded to an integer) and count the frequencies $(\hat{n}_1,\ldots,\hat{n}_5)$ where $\hat{n}_j$ denotes the frequency of all $j$-star-ratings. We only draw five estimators because the RETRAIN study has only five re-ratings and we would like to stay as close as possible to the real data. To cope with the randomness that arises by considering only five draws, we just repeat this procedure a thousand times. The so obtained frequency vectors can be compared to the original $(n_1,\ldots,n_5)$ from our study and $p_{0}$ emerges as the relative frequency of matchings.
For $p_{c}$ we basically follow the same procedure as for $p_{0}$, except that the five estimators are not drawn from the user model but from a uniform distribution.\\[.25ex]
\item[Jensen–Shannon divergence:] This metric is in line with the spirit of the Bayesian brain paradigm since it assumes the user feedback to have a full probability density rather than considering only five values. Each user-item-pair is associated to a normal distribution obtained by ML-fitting on the corresponding re-ratings. For the model-based feedback, we compute $10^6$ estimators $\varphi(\varrho(\xi))$ and also apply ML-fitting. Hence, we yield the probability distributions $P_{\text{model}}$ and $P_{\text{real}}$ for which we compute the Jensen–Shannon-Divergence (JSD)
\begin{eqnarray}
\operatorname{JSD}(P_{\text{model}}| P_{\text{real}})  
&=& \frac{1}{2} D_\mathrm{KL}(P_{\text{model}}| M) \nonumber \\
&&  + \frac{1}{2} D_\mathrm{KL}(P_{\text{real}}| M),
\end{eqnarray}
where $D_\mathrm{KL}(P_1|P_2)=\sum _{i}P_1(i)\,\log_2 (P_1(i)/P_2(i))$ denotes the Kullback-Leibler-Divergence and  $M=\frac {1}{2}(P_{model}+P_{\text{real}})$. 
Since we use the base $2$ logarithm, the JSD yields the boundaries 
\begin{equation}
0 \leq \operatorname{JSD} \leq 2\log(2)  \quad\text{or}\quad 0 \leq \tfrac{\operatorname{JSD}}{2\log(2)} \leq 1.
\end{equation}
The inequality on the right provides a normed metric to evaluate the disparity of probability distributions.\\[.25ex]
\end{description}

\begin{figure*}[b]
    \centering
    \includegraphics[width=\linewidth]{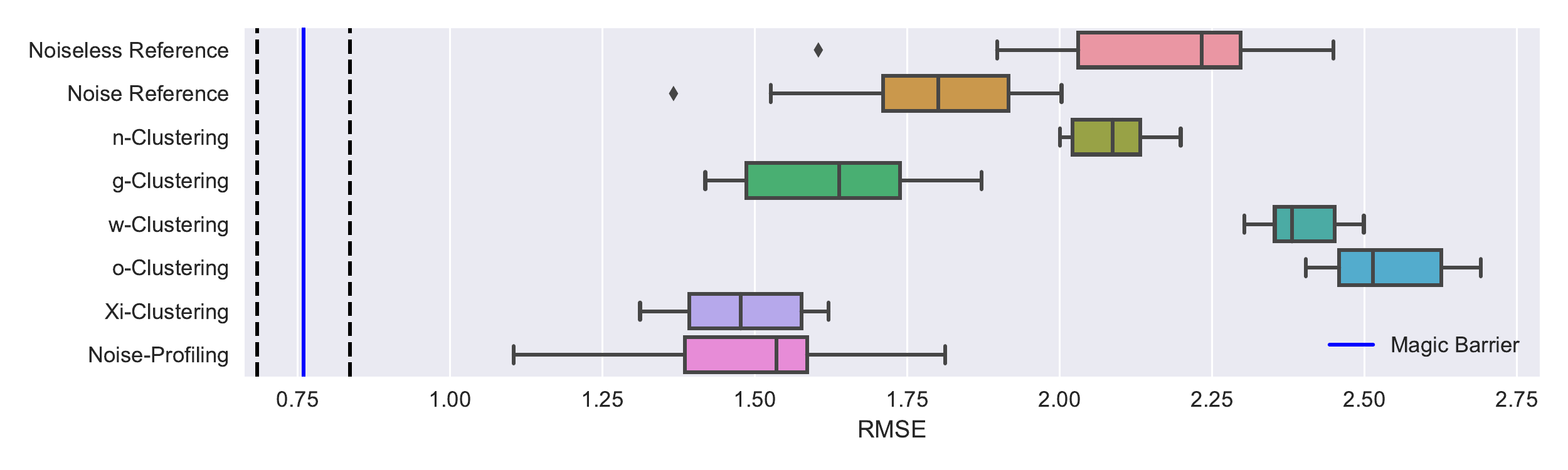}
    \caption{Feedback distributions obtained from different decoder functions for the cognition vector
     $\xi^\mathcal{B}=(100,1,1,5,3)$.}
    \label{fig:Benefits}
	\vspace{-2ex}
\end{figure*}

For a perfect user model, one expects that only a single combination of cognition vector and decoder will make the disparity metric $d$ vanish and that all other combinations will maximize $d$. Therefore, we will not only consider the metric scores themselves, but also their ambiguity. 
The results show that each decoder function is able to fit constant users when using a sufficiently high frequency gain or sufficiently small tuning curve widths. The average ambiguity is 300, i.e. for about 300 cognition vectors we yield the same minimal metric score. Nevertheless, with the lowest-energy-principle from Eq. \ref{eq:Energy}, we can select a single vector and represent users with constant behaviour. However, it becomes clear that the strength of our neuroscientific user models is clearly in the modelling of human uncertainty. Therefore, in the following analyses, we will only consider those users who had provided unreliable feedback. \\[.25ex]

\begin{figure}[t]
    \centering
    \includegraphics[width=\linewidth]{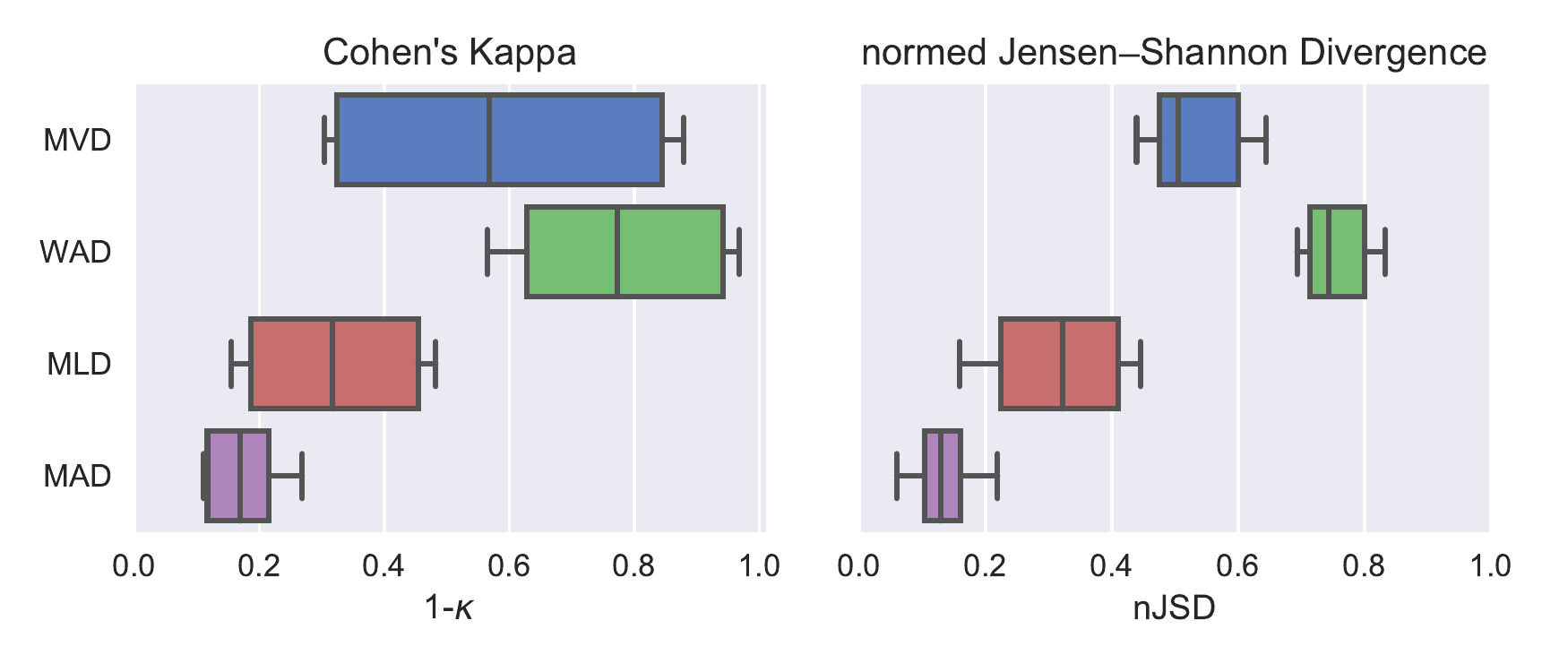}
    \caption{Metric scores for the best fitting cognition vectors aggregated from all user-item-pairs.}
    \label{fig:Quality}
	\vspace{-2ex}
\end{figure}

For noisy users, the mean ambiguity is 5, i.e. only five out of  $10^9$ cognition vectors lead to the same metric minimisation. In Fig. \ref{fig:Quality} we can see the distribution of metric scores for best fitting cognition vectors.
For the descriptive evaluation with Cohen's Kappa, we see that the MVD and WAD perform very poorly.
Sometimes there are user-item-pairs whose best fit is $1-\kappa = 0.8$ or even higher. The best decoders are the MLD and the MAD. It is also noteworthy that there are major overlaps when using this metric. This means that there is ambiguity for the decoder function as well. For example, a large proportion of scores for the MLD can also be achieved by the MVD. Moreover, considering the whiskers of the MAD, half of the scores can also be formed by the MLD. 
Nonetheless, first rankings in model quality can be anticipated.
For the normed Jensen-Shannon divergence, this ranking can be verified. In addition, we can notice the increased amount of information when using distributions rather than samples with five draws. So, the scattering of metric scores is much smaller. In summary, the maximum a posteriori decoder can be mentioned as the best decoder function leading to feedback distributions modelling  reality with high quality. Therefore, we will confine ourselves to this decoder function for further elaborations.

\subsection*{Information Extraction}
Finally, we discuss how adaptive systems benefit from these new information, provided by a high-dimensional neural space. In doing so, we consider collaborative filtering (CF) in its simplest form: User-item-pairs with corresponding product ratings are clustered into user groups in order to recommend new products on the basis of group popularity. In order to compute a reference for further comparisons, we use a simple k-means approach to find clusters within the samples
\begin{equation}
\text{Sample}_t = \{(u,i,\text{rating}_t)\colon u=1,\ldots,67 \, , \, i=1,\ldots,5\}
\end{equation}
separately for each rating trial $t=1,\ldots,5$.
In each sample, we randomly select 30\% of the users in each cluster group to delete their ratings for the fifths item (testing-users). We use the mean rating from the remaining 70\% of users (learning-users) within each group as the specific group predictor for item 5. This predictor is then compared to the original prediction of the testing-users by means of the RMSE. 
In this way, we get an RMSE score for each rating trial, and if we repeat the randomised selection of testing-users five times, we get 25 scores that form a distribution. This approach will be referred to as \textit{noiseless reference}.

Since this approach does not consider any uncertainty information, we need a second reference to fill this gap.
For this purpose, we primarily proceed like above. The only difference is that we execute clustering on the union $\cup_{t=1}^5 \text{Sample}_t$, i.e. we allow copies of user-item-pairs but with different ratings. Our cluster groups will therefore be much larger and means (predictions) more accurate. Additionally, we do not compare the predictions with ratings of a particular trial, but with the mean rating aggregated from all rating trials. This stochastic approach will be referred to as \textit{noisy reference}.\\[.125ex]

In contrast, we introduce the following methods, which are based on the additional information of the nPPC user model:
\begin{description}[leftmargin=6mm]
\item[$\xi$-Clustering:] We associate $\xi_{u,i}$ to each user-item-pair and use k-means on the neural space $\{(u,i,n,g,w,o,s)\}$. We then proceed with selecting testing-users and learning-users, just as for the references above. 
Due to the higher dimensional space, user groups may be much more differentiated and more appropriate for testing-user predictions.

\item[Subspace-Clustering:] Here, we associate $\xi_{u,i}$ to each user-item-pair and use k-means on the neural subspaces $\{(u,i,n)\}$ (denoted as $n$-Clustering), $\{(u,i,g)\}$ ($g$-Clustering), $\{(u,i,w)\}$ ($w$-Clustering), $\{(u,i,o)\}$ ($o$-Clustering). We then proceed as above.

\item[Noise-Profiling:] We associate $\xi_{u,i}$ to each user-item-pair and aggregate by users to yield sets 
$S_u:= \left\lbrace \xi_{u,i} \colon i=1, \ldots , 4 \right\rbrace$ in which we consider only the first four items. We simply  calculate the mean cognition vector $\bar{\xi}_{u,i} = (\bar{n},\bar{g},\bar{w},\bar{o},s)$ for each $S_u$, where $s$ is left arbitrary. Subsequently, we chose $s$ so that the variance of the model-based feedback distribution $(\varphi\circ\mathcal{R})(\bar{\xi}_{u,i})$ is as close as possible to the user's average variance gathered from the rating distributions of the remaining four items.
\end{description}

The results are depicted in Fig. \ref{fig:Benefits}. First of all, it has to be noted that the variances of the RMSE distributions are relatively large, which is due to the size of our data record. As a visualisation of the RMSE's offset (which emerges for uncertain user data), we additionally calculated the magic barrier as proposed by \cite{MagicBarrier} together with its 95\%-confidence interval. 
We can see that the noisy reference operates much better than the noiseless reference. Moreover, we see that the w-clustering and the o-clustering behave much worse than both references. This can be explained by the fact that clustering according to user ratings for predicting other ratings can be regarded as sensible since there is a causality. In contrast, the tuning curve width as well as the offset are not causally related to the user ratings. Actually, one would expect the same for the n-clustering and g-clustering respectively. However, the n-clustering performs a little better than the noiseless reference, although both distributions have a complete intersection. The results for the g-clustering is quite surprising since it outperforms the noisy reference. We explain this by a latent causal dependency between a particular rating and neural frequency. As previously mentioned, information is primarily encoded in terms of frequencies within the human brain. Therefore, frequencies might encode ratings and uncertainty simultaneously. 
For the $\xi$-clustering as well as for the noise-profiling we can certify an excellent performance result. However, there are some  overlaps between all these distributions. For example, the left whisker of the noisy reference reaches the third quartile of the noise-profiling approach. Hence, the noise-profiling does perform doubtlessly better for only 75\% of the data whereas the superiority for the other 25\% is associated with a certain doubt. 
Nevertheless, the success of the neuroscientific user models against this stochastic uncertainty model is quite clear, although one should also consider that we have only investigated a very simple approach of collaborative filtering. A focused investigation of more complicated and more sophisticated techniques is therefore needed and will be done in future research.

\section{Discussion}

In this contribution we have broken with the view that user noise or human uncertainty is something undesirable that only causes trouble in the evaluation of adaptive systems. We explicitly permitted this human property and developed a user model using noisy probabilistic population codes (nPPCs) to reveal and exploit the inherent information. For this purpose, we formulated three research questions at the beginning. 

The first question was about how a possible user model could look like that takes into account human uncertainty.
For this we consider a population of neurons whose noisy tuning curves are equidistantly allocated over an estimation scale (e.g. rating scale). These tuning curves can be adjusted by various parameters, which we represent in a so-called cognition vector.
By this preliminary fixing, the population provides an unreliable response to a stimulus (e.g. a choice of a particular user rating), which can be converted into a real answer through  decoder functions. By means of two disparity metrics we can find a cognition vector together with a decoder function for each user-item-pair so that measured feedback distributions can be reproduced.

The second research question focused on possible solutions for making the information available to adaptive systems.
For this we have chosen the example of collaborative filtering. The simplest and most efficient method is the clustering of user groups based on the neural parameters. These represent a higher-dimensional vector space than normally yielded by clustering for  ratings only. The first results are very promising. We also revealed that the neuroscientific user models outperform a mere statistical model for representing uncertainty.

The third research question referred to the possible benefits of this novel paradigm in general.
Every personalisation engine and every recommender system has the goal of being able to map the human being as accurately as possible. A knowledge of the nature of man, together with his or her peculiarities, is hence crucial.
The theory of nPPCs is currently a much-debated theory and is considered by many neuroscientists to be an adequate model of human  decision-making which is very close to real structures. The Bayesian brain paradigm is always seen in a prominent role and has been verified many times in neurological experiments. Such a theory about human cognitions is hence a decisive possibility to reach for the goal of adaptive systems and to map human beings according to their very nature.
But also an epistemological component is delivered by this contribution. The nPPCs, which have so far only been investigated for sensory perception, have been used for investigating cognitions for the very first time The performance of these models on decision-making is thus a very good result for theoretic neuroscience as well.

\paragraph*{Future Research}
In this article, we only examined bell-shaped tuning curves. However, sigmoid-shaped tuning curves were also frequently measured in vivo. Further investigations of these shapes with respect to our user model are therefore absolutely necessary.
For example, initial results show that the population activity for sigmoid-shaped tuning curves forms convex and concave functions, which are the basis for the utility theory, i.e. the most widely used theory for human decision making in the field of economics.
However, the present model still needs to be extended by many factors and correlates. For example, there might be dependencies between the cognition vector and the evaluation duration, the testimonial length, the revaluations of a given rating, but also the weather, acute stress and emotional states are possible candidates for biasing factors.
Further research will also focus on accelerating the classification approach as brute force is very slow and expensive. For example, the runtime of the classification of $1\, 675$ user-item-pairs for the MAD was about 6 days using multi-processing on 400 CPUs with 2TB RAM of our university's high performance cluster. 


\vfill

\subsection*{Acknowledgements}
Computational support and infrastructure was provided by the Centre for Information and Media Technology (ZIM) at the University of Duesseldorf (Germany).

\bibliographystyle{ACM-Reference-Format}
\bibliography{umap18} 
\balance

\end{document}